\documentclass[aps,prx,twocolumn,superscriptaddress,10pt]{revtex4-2}

\usepackage[scaled=0.95]{helvet}
\usepackage{sansmath}
\usepackage{etoolbox}

\usepackage{amsmath}
\usepackage{amssymb}
\usepackage{bm}
\usepackage[version=4]{mhchem}
\usepackage{textcomp}
\usepackage{gensymb}

\DeclareUnicodeCharacter{2212}{\textendash}

\usepackage{array}[=2023-11-01]
\usepackage{graphicx}
\usepackage{overpic}
\usepackage{booktabs}
\usepackage[table,dvipsnames]{xcolor}
\usepackage{colortbl}
\usepackage{tabularx}
\usepackage{dcolumn}
\usepackage{blkarray}
\usepackage{siunitx}

\makeatletter
\AddToHook{begindocument/end}{%
  \let\insert@pcolumn\insert@column
}
\makeatother

\usepackage{subfiles}
\usepackage{xspace}
\usepackage{comment}
\usepackage{wasysym}
\usepackage{placeins}
\usepackage{ragged2e}
\usepackage[noindentafter]{titlesec}
\usepackage{fancyhdr}

\usepackage[
  colorlinks=true,
  pdfborder={0 0 0},
  linkcolor=blue,
  citecolor=red,
  filecolor=yellow,
  urlcolor=blue,
  bookmarks,
  pdfauthor={}
]{hyperref}

\usepackage[capitalise]{cleveref}
\usepackage{pdfpages}
\makeatletter
\AtBeginDocument{\let\LS@rot\@undefined}
\makeatother
\newcommand{\tablesize}{%
  \fontsize{9}{10}\selectfont
}

\newcommand{\tabletitlesize}{%
  \fontsize{11}{12}\selectfont
}

\newcommand{\tablecaptionsize}{%
  \fontsize{7}{8}\selectfont
}

\newcommand{\tablefont}{%
  \sffamily
  \sansmath
  \tablesize
}

\AtBeginEnvironment{table}{\tablefont}
\AtBeginEnvironment{table*}{\tablefont}
\AtBeginEnvironment{tabular}{\tablefont}
\AtBeginEnvironment{tabularx}{\tablefont}

\newcolumntype{Y}{>{\centering\arraybackslash}X}
\newcolumntype{L}{>{$}l<{$}}
\newcolumntype{R}{>{$}r<{$}}
\newcolumntype{C}{>{$}c<{$}}

\titleformat{\section}
  {\normalfont\sffamily\large\bfseries}
  {\thesection}
  {0pt}
  {}

\titleformat{\subsection}
  {\normalfont\sffamily\bfseries}
  {\thesubsection}
  {0pt}
  {}

\titleformat{\subsubsection}
  {\normalfont\sffamily\bfseries}
  {\thesubsubsection}
  {0pt}
  {}

\titlespacing{\section}
  {0cm}
  {0.7cm}
  {0.01cm}

\titlespacing{\subsection}
  {0cm}
  {0.45cm}
  {0cm}

\titlespacing{\subsubsection}
  {0cm}
  {0.35cm}
  {0cm}

\makeatletter

\renewcommand{\figurename}{Fig.}
\renewcommand{\tablename}{Table}
\renewcommand{\thefigure}{\arabic{figure}}
\renewcommand{\thetable}{\arabic{table}}

\newcommand{\captionbar}{%
  \rule[-0.5ex]{0.8pt}{2.35ex}%
}

\renewcommand{\fnum@figure}{%
  \textbf{\figurename~\thefigure~}%
  \captionbar
}

\renewcommand{\fnum@table}{%
  \tablefont
  \textbf{\tablename~\thetable~}%
  \captionbar
}

\long\def\@makecaption#1#2{%
  \par
  \vskip\abovecaptionskip
  \begingroup
    \normalfont
    \fontsize{8}{9}\selectfont
    \noindent
    \vbox{%
      \hsize=\hsize
      \leftskip=0pt
      \rightskip=0pt
      \parfillskip=0pt plus 1fil
      \noindent #1\ #2\par
    }%
  \endgroup
  \vskip\belowcaptionskip
}

\makeatother

% ...
\newcommand{\tabletitle}[2]{%
  \refstepcounter{table}%
  \label{#2}%
  \par
  \noindent
  \makebox[\linewidth][l]{%
    \sffamily
    \sansmath
    \tabletitlesize
    \textbf{Table~\thetable~}%
    \captionbar
    \hspace{0.45em}%
    #1%
  }%
  \par
  \vspace{4pt}%
}

\newcommand{\tablecaption}[1]{%
  \par
  \vspace{4pt}%
  \noindent
  \parbox{\linewidth}{%
    \sffamily
    \sansmath
    \tablecaptionsize
    \RaggedRight
    #1\par
  }%
}

\newcommand{\authorfont}{%
  \sffamily
  \bfseries
  \normalsize
}

\newcommand{\affiliationfont}{%
  \sffamily
  \normalfont
  \fontsize{8}{10}\selectfont
}

\newcommand{\articledatefont}{%
  \sffamily
  \normalfont
  \fontsize{8}{10}\selectfont
}

\newcommand{\abstractfont}{%
  \sffamily
  \normalfont
  \fontsize{10.5}{13}\selectfont
}

\AddToHook{begindocument/end}{%
  \makeatletter

  \def\frontmatter@title@format{%
    \sffamily
    \bfseries
    \raggedright
    \fontsize{22}{25}\selectfont
  }

  \def\frontmatter@title@below{%
    \par
    \vspace{0.5em}%
    \noindent
    \rule{\textwidth}{0.6pt}%
    \par
    \vspace{0.7em}%
  }

  \def\frontmatter@authorformat{%
    \authorfont
    \centering
    \parindent=0pt
    \parskip=0pt
  }

  \def\frontmatter@above@affiliation@script{%
    \centering
    \addvspace{5pt}%
  }

  \def\frontmatter@affiliationfont{%
    \affiliationfont
    \centering
    \parindent=0pt
    \parskip=0pt
    \hyphenpenalty=10000
    \exhyphenpenalty=10000
    \emergencystretch=2em
  }

  \def\frontmatter@RRAP@format{%
    \articledatefont
    \centering
  }

  \def\frontmatter@abstractfont{%
    \abstractfont
    \RaggedRight
    \parindent=0pt
    \parskip=0pt
    \adjust@abstractwidth
  }

  \def\frontmatter@abstractwidth{%
    0.82\textwidth
  }

  \def\adjust@abstractwidth{%
    \dimen@=\textwidth
    \advance\dimen@-\frontmatter@abstractwidth
    \leftskip=0pt
    \rightskip=\dimen@ plus 2em
    \@totalleftmargin=0pt
  }

  \def\frontmatter@preabstractspace{%
    0pt
  }

  \def\frontmatter@postabstractspace{%
    2.2\baselineskip
  }

  \makeatother
}

\let\originaldate\date

\renewcommand{\date}[1]{%
  \originaldate{{\articledatefont #1}}%
}

\makeatletter
\apptocmd{\frontmatter@author@produce@script}{%
  \par
  \vspace{0.7em}%
  \noindent
  \rule{\textwidth}{0.6pt}%
  \par
  \vspace{0.8em}%
}{}{}
\makeatother

\fancypagestyle{plain}{%
  \fancyhf{}
  \fancyfoot[R]{%
    \sffamily
    \bfseries
    \footnotesize
    \thepage
  }

}

\AddToHook{begindocument/end}{%
}

\begin{document}

\title{Quetzalcoatlite as a Disorder-Free Platform for Chiral Magnetism and Frustration}

\author{Aleksandar Razpopov}
\email{razpopov@itp.uni-frankfurt.de}
\affiliation{Institut f\"ur Theoretische Physik, Goethe-Universit\"at Frankfurt, 60438 Frankfurt am Main, Germany}

\author{P. Peter Stavropoulos}
\email{panagiotis@itp.uni-frankfurt.de}
\affiliation{Institut f\"ur Theoretische Physik, Goethe-Universit\"at Frankfurt, 60438 Frankfurt am Main, Germany}

\author{Felix Flicker}
\email{flicker@physics.org}
\affiliation{School of Physics, University of Bristol, Bristol, BS8 1TL, UK}

\author{Michael R. Norman}
\email{norman@anl.gov}
\affiliation{Materials Science Division, Argonne National Laboratory, Lemont, IL 60439, USA}

\author{Roser Valent\'i}
\email{valenti@itp.uni-frankfurt.de}
\affiliation{Institut f\"ur Theoretische Physik, Goethe-Universit\"at Frankfurt, 60438 Frankfurt am Main, Germany}

\begin{abstract}
\textbf{Abstract:} The natural mineral quetzalcoatlite \ce{Zn_6Cu_3(TeO_6)_2(OH)_6 \cdot (Ag_xPb_yCl_{x+2y})} is a structurally ideal kagome magnet, providing a platform for exploring the interplay of geometric frustration, chirality, and tunability in a disorder-free framework. Here, we present the (first) comprehensive \textit{ab initio} study of its electronic and magnetic properties. The electronic structure is dominated by localized half-filled Cu $d_{x^2-y^2}$ orbitals that become insulating through electronic correlations. Mapping the low-energy physics onto a Heisenberg model reveals that the magnetism is governed primarily by two exchange interactions: a nearest-neighbor intralayer kagome coupling and a next-nearest-neighbor interlayer coupling. Their competition stabilizes an unconventional three-dimensional chiral magnetic state. Each kagome layer hosts a $\sqrt{3}\times\sqrt{3}$ order, while adjacent layers are rotated by $60^\circ$, producing a right-handed spiral along the crystallographic $c$-axis. This intrinsic chiral order emerges naturally from the crystal structure and magnetic interactions, establishing quetzalcoatlite as a distinctive realization of chiral magnetism on a perfect kagome lattice. At the same time, the small energy scale of the exchange interactions places the material close to competing magnetic regimes, suggesting that moderate pressure, chemical substitution, or structural modifications may strongly enhance frustration, suppress long-range order, and potentially drive the system toward a quantum spin-liquid state.
\end{abstract}
\date{\today}
\maketitle

\phantomsection
\section{Introduction}

Magnetic frustration arises when interactions between local moments cannot be simultaneously satisfied, preventing the system from minimizing all pairwise interaction energies at once \cite{RamirezARMR1994,MoessnerPT2006,BalentsNature2010}. As a result, frustrated magnets can exhibit a variety of unconventional magnetic states, including chiral magnetic order \cite{KawamuraJPCondMat1998,KawamuraCanada2001}, or relieve frustration by forming highly entangled ground states such as quantum spin liquids \cite{BalentsNature2010,SavaryRepProgP2016,BroholmSci2020}. A prototypical example is the antiferromagnetic (AFM) triangular lattice, where the lattice geometry precludes the simultaneous satisfaction of all nearest-neighbor AFM interactions \cite{WannierPR1950,HoutappelPhysica1950,AndersonMatResB1973}. This realization has motivated an extensive search for materials that naturally host frustrated lattice geometries \cite{RamirezARMR1994,BalentsNature2010,ZhouRMP2017,BroholmSci2020}, with kagome-lattice compounds emerging as one of the most prominent platforms for exploring frustration-induced quantum phases \cite{MendelsJPSJ2010,NormanRMP2016,ZhouRMP2017,BroholmSci2020}.

One of the earliest and most notable realizations of an effective kagome lattice is the copper mineral herbertsmithite \cite{ShoresJAChemS2005,HeltonPRL2007,olariu200817,NormanRMP2016}. In this material, magnetic \ce{Cu^{2+}} ions form two-dimensional kagome layers that are separated by non-magnetic \ce{Zn^{2+}} ions. Extensive experimental studies have found no evidence of long-range magnetic order down to the lowest measured temperatures \cite{HeltonPRL2007,deVriesPRL2008,deVriesPRL2009,ChengPRB2011}, despite a large negative Curie--Weiss temperature \cite{BertPRB2007,HeltonPRL2007} indicating strong antiferromagnetic interactions. A major experimental challenge, however, is the intrinsic \ce{Cu/Zn} antisite disorder \cite{FreedmanJACS2010,deVriesPRL2008,deVriesPRB2012,KozlenkoPRL2012,HanPRB2016,kremer2925chemical}, which complicates the interpretation of experimental observations and has motivated an ongoing search for polymorphs and related kagome compounds in which this disorder is minimized or eliminated.

%Quetzalcoatlite is another copper mineral that has been known in the mineralogy community for many years \cite{burns2000quetzalcoatlite}. Unlike herbertsmithite, where \ce{Cu/Zn} antisite disorder due to the same local envirolement of Zu and Cu strongly complicates the interpretation of experimental results \cite{FreedmanJACS2010,deVriesPRL2008,deVriesPRB2012,KozlenkoPRL2012,HanPRB2016}, and similarly affects other kagome minerals such as kapellasite and members of the \ce{Zn}-substituted barlowite family \cite{ColmanChemM2008,JansonPRL2008,FaakPRL2012,KermarrecPRB2014,PuphalPRM2018,SmahaPRM2020,BrendanNPJQM2020}, such disorder is expected to be strongly suppressed in quetzalcoatlite due to the distinct octahedral \ce{Cu} and tetrahedral \ce{Zn} coordination environments.
%Quetzalcoatlite is another copper mineral that has long been known in the mineralogy community \cite{burns2000quetzalcoatlite}.
%In contrast to herbertsmithite, where \ce{Cu/Zn} antisite disorder arising from similar octahedral local coordination environments complicates the interpretation of experimental results \cite{FreedmanJACS2010,deVriesPRL2008,deVriesPRB2012,KozlenkoPRL2012,HanPRB2016}, and other kagome minerals such as kapellasite and \ce{Zn}-substituted barlowite, which exhibit similar disorder \cite{ColmanChemM2008,JansonPRL2008,FaakPRL2012,KermarrecPRB2014,PuphalPRM2018,SmahaPRM2020,BrendanNPJQM2020}, such disorder is expected to be suppressed in quetzalcoatlite due to the distinct octahedral \ce{Cu} and tetrahedral \ce{Zn} coordination environments.
Quetzalcoatlite is another copper mineral that has long been known in the mineralogy community \cite{burns2000quetzalcoatlite}.
In herbertsmithite, \ce{Cu/Zn} antisite disorder arising from similar octahedral coordination environments complicates the interpretation of experimental results \cite{FreedmanJACS2010,deVriesPRL2008,deVriesPRB2012,KozlenkoPRL2012,HanPRB2016}.
Similar disorder also affects other kagome minerals, such as kapellasite and \ce{Zn}-substituted barlowite \cite{ColmanChemM2008,JansonPRL2008,FaakPRL2012,KermarrecPRB2014,PuphalPRM2018,SmahaPRM2020,BrendanNPJQM2020}.
In contrast, such disorder is expected to be strongly suppressed in quetzalcoatlite due to the distinct octahedral \ce{Cu} and tetrahedral \ce{Zn} local environments.

Furthermore, while herbertsmithite exhibits substantial \ce{Cu-O-Cu} superexchange, resulting in a relatively large nearest-neighbor exchange interaction ($\sim 17\text{meV}$) \cite{HeltonPRL2007,JeschkePRB2013,NormanRMP2016}, this strong interaction complicates neutron-based investigations \cite{deVriesPRL2009,HanNature2012,NormanRMP2016,SmahaPRB2023}. In quetzalcoatlite, the exchange interactions are expected to be significantly weaker due to the more extended \ce{Cu-O-Te-O-Cu} exchange pathways, which introduce additional buffering through the tellurium environment and lead to longer \ce{Cu-O-O-Cu} paths.
Consequently, quetzalcoatlite is expected to exhibit smaller magnetic energy scales than herbertsmithite. Despite its potential as a clean kagome platform, it remains relatively unexplored, with only a few studies reported to date, including Raman spectroscopy of localized \ce{OH-} vibrational modes \cite{frost2009raman}.

%\mn{You might want to contrast with herbertsmithite in the intro.  HS has a large SE which complicates studying it by neutrons (J is 17 meV).  But because of the Cu-O-Te-O-Cu lattice for QS, exchange constants are significantly reduced.  It should also be stated that we do not expect Cu-Zn interchanges here, since Cu and Zn are in very different crystallographic environments (unlike HS).  Also, for color, you can state that QS is well known in mineralogy community but zero physics work has been done on it.}
%\ar{-kagome lattice, no  expected disorder, compared to y-kapellasite, herbertsmithite }\\
%\ar{-Perfect kagome lattice with NN J, the system exhibits a highly degenerate classical spin liquid ground state~\cite{reimers1993order}}\\
%\ar{-Limited experiments: X-ray experiments~\cite{burns2000quetzalcoatlite} to obtain the crystal structure, and Raman study~\cite{frost2009raman} vibrational modes}
\begin{figure*}
	\centering
	\begin{overpic}[width=1.0\linewidth,percent,grid=false,tics=1]{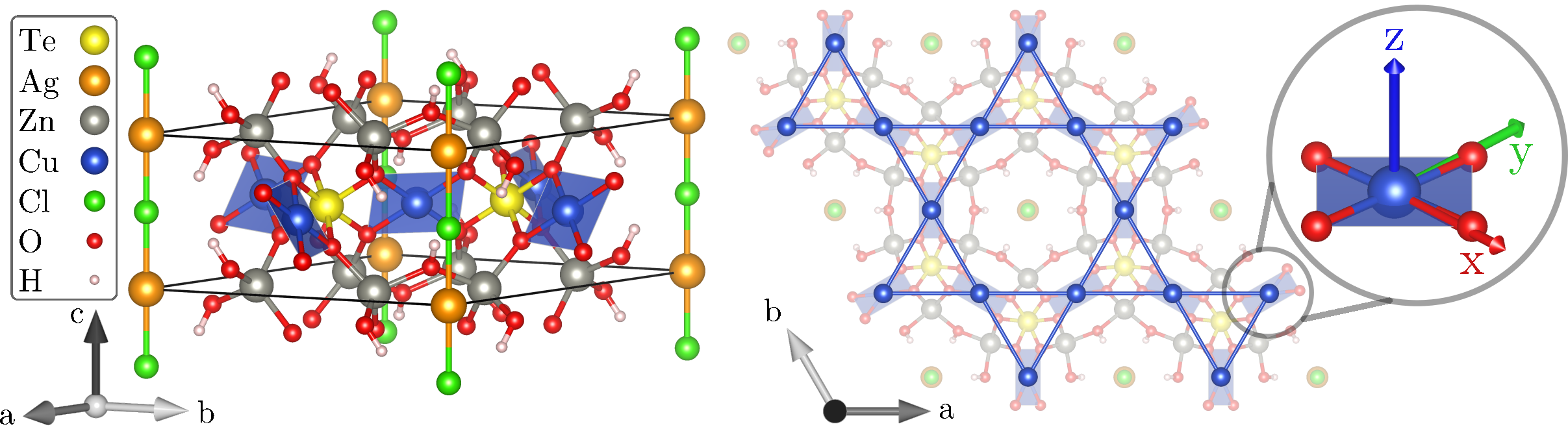}
		\put(0.5,27){\textbf{a}}
		\put(47.1,27){\textbf{b}}
		\put(81,27){\textbf{c}}
	\end{overpic}
	\caption{\textbf{Crystal structure of quetzalcoatlite.} The idealized stoichiometry structure ($x=1$ and $y=0$ in \ce{Zn_6Cu_3(TeO_6)_2(OH)_6 \cdot (Ag_xPb_yCl_{x+2y})}) is shown. \textbf{a} Side view of the primitive unit cell. \textbf{b} Top-down view, with the characteristic kagome lattice formed by the magnetic \ce{Cu^2+} sites, decorated by \ce{O} and \ce{OH-} ligands. \textbf{c} Closeup view of the local \ce{Cu} environment, with the local axis $\mathrm{z}$ perpendicular to the ligand plane, and $\mathrm{xy}$ mostly aligned with the oxygens. The crystal structure was visualized using VESTA~\cite{VESTA}.}
	\label{fig:crystalcombined}
\end{figure*}

Motivated by the structurally ideal kagome lattice of quetzalcoatlite, we present in this work the first (to our knowledge) \textit{ab initio} investigation of its electronic and magnetic properties. 
We show that the low-energy electronic structure is dominated by correlated, localized half-filled Cu $d_{x^2-y^2}$ orbitals. By performing a mapping onto an effective Heisenberg model, we identify two leading exchange interactions whose competition stabilizes an intrinsic three-dimensional chiral magnetic ground state, consisting of $\sqrt{3}\times\sqrt{3}$ kagome order and a 60$^\circ$  rotation between adjacent layers along the (c)-axis. The overall exchange energy scale is small, placing quetzalcoatlite in close proximity to competing magnetic phases.

\section{Results}
\subsection{Structural Relaxation}
The structure of quetzalcoatlite, as shown in \cref{fig:crystalcombined}a, determined via X-ray diffraction \cite{burns2000quetzalcoatlite}, is a trigonal system in space group $P\bar{3}1m$ (No. 162), with lattice parameters $a = 10.145$~\AA\ and $c = 4.9925$~\AA.
The \ce{Cu^2+} ions, surrounded by a 4-fold planar oxygen environment, form a perfect kagome lattice within the $a$-$b$ plane, with layers vertically stacked along the crystallographic $c$-axis.
The adjacent kagome planes are separated by intercalated layers of \ce{ZnO2(OH)2} tetrahedrons.
Within the \ce{Cu^2+} kagome network, a \ce{TeO6} polyhedron occupies the center of each kagome triangle, while an interstitial complex \ce{$X$ = Ag_$x$Pb_$y$Cl_{$x+2y$}} resides along the $c$-axis at the center of each hexagon, see top view in \cref{fig:crystalcombined}b.
While the naturally occurring mineral exhibit fractional occupancies $x$, $y$, \textit{ab initio} investigations require a well-defined, periodic crystal structure.
To facilitate first-principles modeling and reduce computational cost, an ideal stoichiometric setting of $x=1$ and $y=0$ is used, thereby omitting any \ce{Pb} mix-in, establishing a pristine cell of \ce{Zn6Cu3(TeO6)2(OH)6AgCl}.
%The \textit{ab initio} investigation requires a well-defined crystal structure,
%while natural mineral specimens display complex fractional occupancies of the Cl site involving $X = \mathrm{Ag}_x\mathrm{Pb}_y\mathrm{Cl}_{x+2y}$ along the $c$-axis interstitial space.
 %We bypass this fractional occupation to facilitate precise first-principles modeling, and reduce the computational cost, by assuming an idealized stoichiometric structure, with $x=1$ and $y=0$, thus
 %establish a unit cell with the chemical formula $\mathrm{Zn_6Cu_3(TeO_6)_2(OH)_6AgCl}$, omitting any $\mathrm{Pb}$ mixing.
Considering the large distance of the complex $X$ has from the magnetic \ce{Cu} sites, magnetic exchange  should be minimally affected, justifying the approximation.

 Due to the X-ray insensitivity to the hydrogen positions, these were obtained by symmetry consideration and \textit{ab initio} position optimization while keeping all other crystal parameters fixed
 , see \hyperref[sec:methods]{Methods}, and Supplemental Information for more details.
 The resulting crystal structure is summarized in~\cref{tab:quetzalcoatlite_positions}.
 We also performed an unconstrained full structural relaxation.
 The volume increased by approximately 4.5\%, but showed negligible impact on the electronic properties.
 Consequently, to maintain fidelity to the empirical data, we performed all subsequent calculations using the experimental lattice parameters, with only the hydrogen position determined from the relaxation.

\begin{table}
\centering

\tabletitle{\textbf{Fractional atomic coordinates}}{tab:quetzalcoatlite_positions}

\begingroup
\tablefont
\setlength{\tabcolsep}{0pt}
% Keep row backgrounds flush with the horizontal rules.
\setlength{\aboverulesep}{0pt}
\setlength{\belowrulesep}{0pt}
\renewcommand{\arraystretch}{1.15}

\begin{tabular}{
		@{}
		w{l}{0.22\columnwidth}
		w{c}{0.26\columnwidth}
		w{c}{0.26\columnwidth}
		w{c}{0.26\columnwidth}
		@{}
}
\toprule
\rowcolor{gray!30}
\textbf{\hspace{3pt}Atom}
& $\mathbf{x}$
& $\mathbf{y}$
& $\mathbf{z}$ \\
\midrule

\hspace{3pt}Te
& $1/3$
& $2/3$
& $1/2$ \\
\specialrule{\heavyrulewidth}{0pt}{0pt}

\rowcolor{gray!10}
\hspace{3pt}Zn
& $0.20370$
& $0.79630$
& $0$ \\
\specialrule{\heavyrulewidth}{0pt}{0pt}

\hspace{3pt}Cu
& $1/2$
& $0$
& $1/2$ \\
\specialrule{\heavyrulewidth}{0pt}{0pt}

\rowcolor{gray!10}
\hspace{3pt}O(1)
& $0.35080$
& $0.51900$
& $0.28500$ \\
\specialrule{\heavyrulewidth}{0pt}{0pt}

\hspace{3pt}O(2)
& $0.27720$
& $0$
& $0.15800$ \\
\specialrule{\heavyrulewidth}{0pt}{0pt}

\rowcolor{gray!10}
\hspace{3pt}H
& $0.22201$
& $0$
& $0.31758$ \\
\specialrule{\heavyrulewidth}{0pt}{0pt}

\hspace{3pt}Ag
& $0$
& $0$
& $0$ \\
\specialrule{\heavyrulewidth}{0pt}{0pt}

\rowcolor{gray!10}
\hspace{3pt}Cl
& $0$
& $0$
& $1/2$ \\
\bottomrule

\end{tabular}

\endgroup

\tablecaption{Fractional atomic coordinates of the stoichiometricaly idealized quetzalcoatlite structure in \cref{fig:crystalcombined}, found from \textit{ab initio} relaxations of the hydrogen position, see \hyperref[sec:methods]{Methods}.}

\end{table}

\subsection{Electronic Properties}
\begin{figure}
	\centering
	\begin{overpic}[width=1.00\linewidth,percent,grid=false,tics=4]{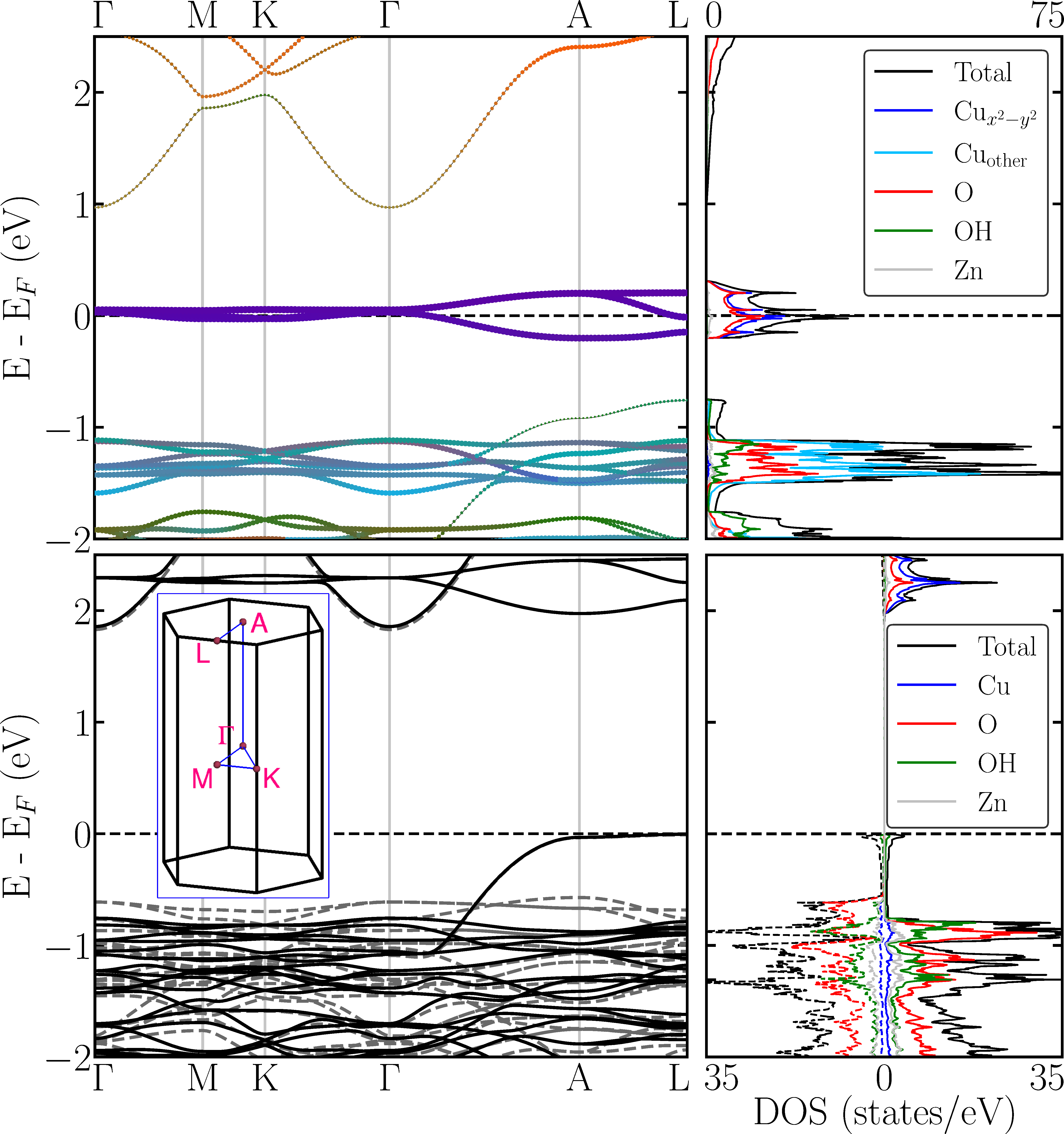}
		\put(0,97.5){\sffamily\textbf{a}}
		\put(0,48){\sffamily\textbf{b}}
	\end{overpic}
	\caption{\textbf{Electronic structure of quetzalcoatlite.}
    \textbf{a} Band structure and atom resolved DOS at the GGA level. 
    The superimposed fat bands indicate the respective contributions in the bands. The \ce{Cu} DOS is given in the local coordinates as defined in~\cref{fig:crystalcombined}c, and additionally resolved in the local $x^2-y^2$ contributions. \textbf{b} Band structure and DOS calculated within the GGA$+U$ framework with 
    $U_{\mathrm{eff}} = 6$~eV,
    %$U=6$~eV,
    assuming a polarized FM configuration. Solid and dashed lines represent the spin-up and spin-down channels, respectively. The negligible Cl and Ag  contribution, in this energy window, are not visible.}\label{fig:ggapdoscombined}
\end{figure}

Utilizing the \textit{ab initio} relaxed crystal structure, we evaluate the electronic properties of quetzalcoatlite.
For this analysis, we define the local coordinate system such that the local $\mathrm{x}$ and $\mathrm{y}$ axes point approximately towards the coordinating oxygen atoms and $\mathrm{z}$ is perpendicular to the local oxygen plane, as illustrated in \cref{fig:crystalcombined}c.
The GGA bands and DOS are presented in \cref{fig:ggapdoscombined}a.
The states crossing the Fermi level are dominated by the half-filled \ce{Cu} $d_{x^2-y^2}$ band, which strongly hybridizes with the neighboring \ce{O} $p$ orbitals.
The presence of this half-filled $d_{x^2-y^2}$ band directly supports the assignment of a \ce{Cu^2+} oxidation state, supporting a \ce{Cu} $d^9$ configuration, with a local spin-$1/2$ moment.
The remaining \ce{Cu} $3d$ orbitals, alongside \ce{O} and smaller \ce{OH-} contributions, are approximately $-1$~eV below the Fermi level, separated by an energy gap.
The spectral weight from all other orbital characters only become significant below $-2$~eV, see Supplemental Information.
Notably, the states associated with the \ce{Cl} and \ce{Ag} ions are negligible in the shown energy window and are therefore omitted from the DOS plots for visual clarity.

Including correlations with the GGA$+U$ functional, with $U=6$~eV, drives the system into an insulating state, opening up  a pronounced gap of $2$~eV, see~\cref{fig:ggapdoscombined}b.
The calculated magnetic moment on the \ce{Cu} sites is fully consistent with an $S=1/2$ state, confirming the physical picture of a pristine kagome lattice hosting localized spin-1/2 magnetic moments. To the best of our knowledge, there is no experimental value for the gap, and we set the $U$ value similarly to other \ce{Cu-O} based minerals~\cite{JeschkePRB2013,hering2022phase}, to capture the qualitatively gapped nature of the strongly correlated system.
%%Further choices of U highlighting the robustness of the model with respect of the U value are shown in the Supplement Information. 
Additional values of $U$, demonstrating the robustness of the model with respect to the choice of $U$, are presented in the Supplemental Information.

\subsection{Magnetic Model}
\begin{figure}
	\centering
	\begin{overpic}[width=1.0\linewidth,percent,grid=false,tics=4]{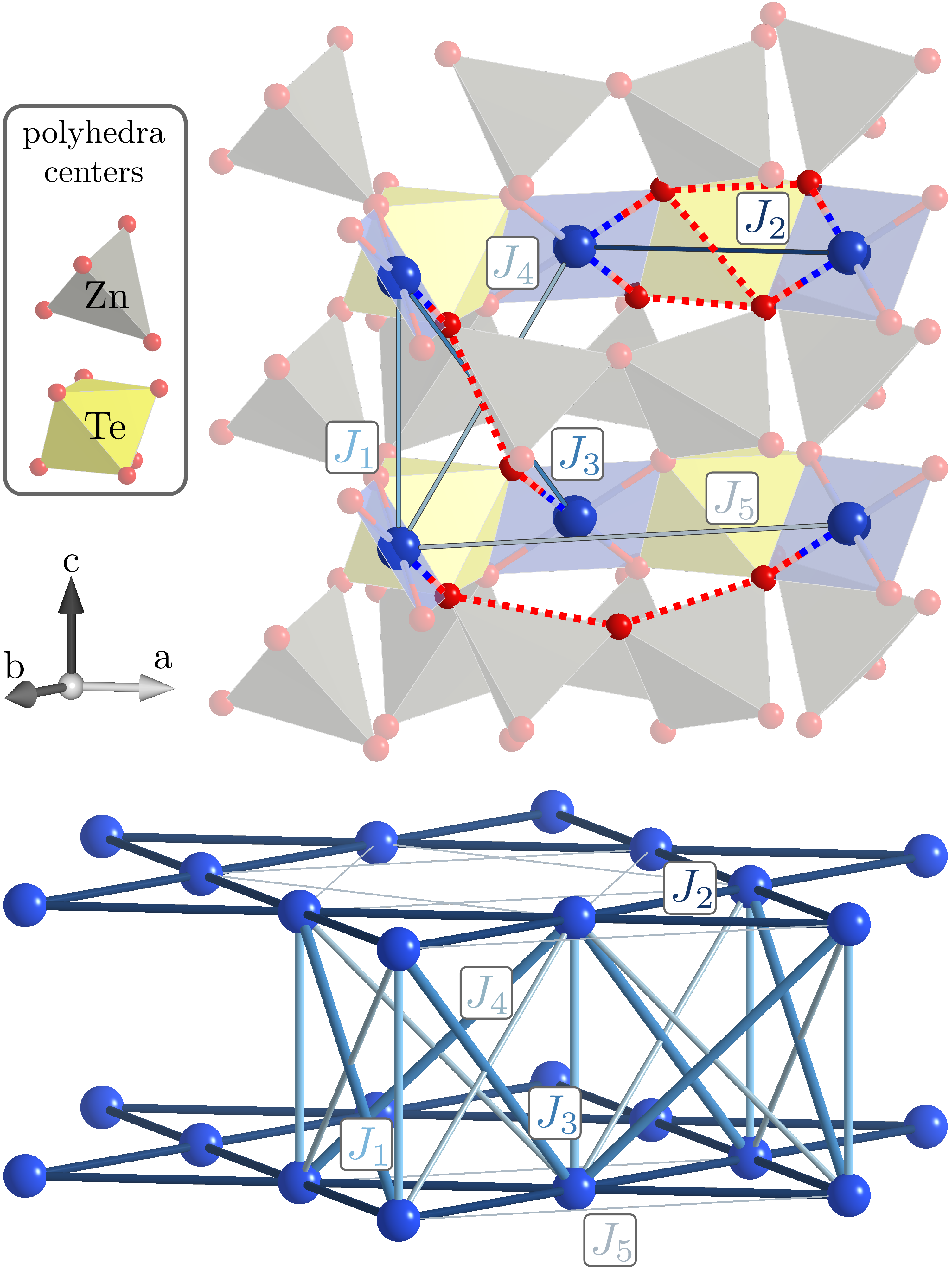}
		\put(0,98){\sffamily\textbf{a}}
		\put(0,37){\sffamily\textbf{b}}
	\end{overpic}
	\caption{\textbf{Exchange microscopic paths and network.} \textbf{a} Cut out of the crystal enviroment, highlighting the superexchange paths for $J_1$ to $J_5$. The contributing \ce{Cu} and \ce{O} sites are highlighted. \textbf{b} The copper magnetic exchange network up to $J_5$, color coded from strongest coupling (dark blue) to weakest (light blue). The crystal structure was visualized using VESTA~\cite{VESTA}}
	\label{fig:j-couplingscombined}
\end{figure}

We construct the magnetic exchange model up to 5\textsuperscript{th}NN
\begin{equation}
    H = \sum_{\alpha=1}^{5}\sum_{\langle i,j \rangle_{\alpha}}\!\! J_{\alpha} \vec{S}_i \!  \cdot \! \vec{S}_j 
\end{equation}
with $J_{\alpha}$ as illustrated in \cref{fig:j-couplingscombined}.
Within the $a$-$b$ plane, $J_2$ bonds form the fundamental kagome network, and $J_5$ corresponds to what would be called a kagome second NN coupling.
Along the $c$-axis, the exchanges $J_1$, $J_3$, and $J_4$ mediate the interlayer magnetic interactions, with $J_1$ being the shortest bond in the structure.
The \ce{Cu-Cu} exchange parameters, bond lengths, and structural assignments, are summarized in \cref{tab:table_jnetwork_quatzo}. 

%Next, we clarify the symmetry relation of  $J_3$ and $J_4$ bonds. These two interlayer exchanges exhibit identical bond length of $7.1173$~\AA but are strictly symmetry-inequivalent.
%This fundamental symmetry breaking is rooted in their distinct local coordination environments.
%The physical manifestation of the symmetry breaking between the $J_3$ and $J_4$ bonds is explicitly shown in \cref{fig:j-couplingscombined}(b).
%ar{No figure about J3 and J4 local enviroment anymore.}
%The $J_3$ superexchange pathway is mediated by two oxygen atoms belonging to the $\mathrm{ZnO}_2\mathrm{(OH)}_2$ tetrahedra, with the $\mathrm{Cu}-\mathrm{O}-\mathrm{O}-\mathrm{Cu}$ subpaths closely aligned to the direct $\mathrm{Cu-Cu}$ bond axis.
%In contrast, no such intermediate oxygen atoms are present along the $J_4$ pathway.
%Consequently, the microscopic orbital overlap and resultant superexchange mechanisms governing $J_3$ and $J_4$ are fundamentally different. 
%The extracted magnetic exchange couplings $J_i$ are reported in \cref{tab:table_jnetwork_quatzo}.

\begin{table}
\centering

\tabletitle{\textbf{Exchange interactions}}{tab:table_jnetwork_quatzo}

\begingroup
\tablefont
\setlength{\tabcolsep}{0pt}
% Keep row backgrounds flush with the horizontal rules.
\setlength{\aboverulesep}{0pt}
\setlength{\belowrulesep}{0pt}
\renewcommand{\arraystretch}{1.15}

\begin{tabular}{
		@{}
		w{c}{0.16\columnwidth}
		S[
		table-format=1.4,
		table-column-width=0.25\columnwidth,
		table-number-alignment=center
		]
		S[
		table-format=+1.3(2),
		table-column-width=0.27\columnwidth,
		table-number-alignment=center
		]
		>{\hspace{3pt}}w{c}{0.32\columnwidth}
		@{}
	}
\toprule
\rowcolor{gray!30}
\textbf{Bond}
& \multicolumn{1}{c}{\textbf{$d_{\ce{Cu-Cu}}$ (\AA)}}
& \multicolumn{1}{c}{\textbf{$J_{\alpha}$ (meV)}}
& \textbf{Assignment} \\
\midrule

$J_1$
& 4.9925
& -0.158(11)
& Interlayer 1NN \\
\specialrule{\heavyrulewidth}{0pt}{0pt}

\rowcolor{gray!10}
$J_2$
& 5.0725
& +1.571(7)
& In-plane 1NN \\
\specialrule{\heavyrulewidth}{0pt}{0pt}

$J_3$
& 7.1173
& +1.432(5)
& Interlayer 2NN \\
\specialrule{\heavyrulewidth}{0pt}{0pt}

\rowcolor{gray!10}
$J_4$
& 7.1173
& -0.037(5)
& Interlayer 2NN \\
\specialrule{\heavyrulewidth}{0pt}{0pt}

$J_5$
& 8.7858
& +0.016(8)
& In-plane 2NN \\
\bottomrule

\end{tabular}
\endgroup

\tablecaption{Bond legths, and Magnetic exchanges couplings $J_{\alpha}$, up to 5\textsuperscript{th}NN, evaluated via the TEMA approach using the GGA$+U$ with an effective Hubbard parameter of $U_{\mathrm{eff}} = 6$~eV. The NN classification in terms of in the kagome plain, and between the kagome planes, is also listed. See \cref{fig:j-couplingscombined} for a visual of the bond structure.}

\end{table}

We utilize the TEMA approach (see \hyperref[sec:methods]{Methods}) to extract the numerical magnetic exchange couplings $J_i$, which are reported in \cref{tab:table_jnetwork_quatzo}.
These values were evaluated at a fixed effective Hubbard parameter of $U_{\mathrm{eff}} = 6$~eV. The magnetic model is heavily dominated by the AFM exchange couplings $J_2$ and $J_3$, with the hierarchy $J_2 > J_3$.
The next most significant coupling is the FM interaction $J_1$, whereas the remaining exchange constants, $J_4$ and $J_5$, are negligibly small.

The relative hierarchy of these exchange couplings can be understood from the local structural geometry of the system, illustrated in \cref{fig:j-couplingscombined}a.
The exchange model is primarily governed by intralayer $J_2$ and interlayer $J_3$, both of which are mediated via $\mathrm{Cu-O-O-Cu}$ superexchange pathways. This is consistent with the projected DOS presented in \cref{fig:ggapdoscombined}a, which reveals that the states near the Fermi level are dominated by the strongly hybridized \ce{Cu} $d_{x^2-y^2}$ and \ce{O} $p$ orbitals.
Consequently, microscopic virtual hopping processes utilizing these overlapping active orbitals yield the largest exchange integrals. Although $J_1$ represents the physically shortest interatomic distance, its corresponding exchange pathway crosses an interstitial gap between the kagome layers, lacking sufficiently near by oxygen ligands to facilitate orbital overlap, resulting in an intrinsically weak coupling. The kagome intraplane 2\textsuperscript{nd}NN coupling, $J_5$, is heavily suppressed due to its large interatomic distance of $\sim 8.8$~\AA, requiring more oxygen ligands in the superexchange path.

Among the interlayer couplings, a critical distinction must be made between $J_3$ and $J_4$.
Although they have identical bond lengths (7.1173~\AA), they are symmetry-inequivalent under the space group.
The symmetry inequivalence is manifest in the distinct local coordination environments. Specifically, the $J_3$ 
superexchange pathway is mediated by two oxygen atoms belonging to the \ce{ZnO4} tetrahedron, 
forming a \ce{Cu-O-O-Cu} 
path closely aligned with the direct \ce{Cu-Cu} 
bond axis.
In contrast, the $J_4$ pathway lacks any short \ce{O-O} subpath, thus suppressing ligand mediated processes. Consequently, despite their equal lengths, the microscopic superexchange mechanisms governing $J_3$ and $J_4$ are fundamentally different, making $J_4$ very small, while $J_3$ is large. This will have important implications on the chirality of the ground state, which we will revisit in the next section.

Comparing the obtained magnetic exchange couplings of quetzalcoatlite to other copper based kagome systems—such as herbertsmithite, kapellasite~\cite{JeschkePRB2013}, and Y-kapellasite~\cite{hering2022phase}—reveals a stark contrast in the underlying exchange mechanisms.
In the latter systems, magnetic interactions are dominated by \ce{Cu-O-Cu} superexchange pathways mediated by a single bridging oxygen.
Quetzalcoatlite, however, features primary exchange paths with at least a \ce{Cu-O-O-Cu} geometry.
The leading-order contributions to the superexchange arise from at least sixth-order perturbation theory, resulting in $J$ couplings being naturally suppressed.
Our computed values reflect this significant magnitude reduction, which should manifest experimentally as a substantially smaller Curie-Weiss temperature.

Finally, we briefly return to our choice of stoichiometry and its impact on the \textit{ab initio} estimates discussed above.
Couplings were extracted using the idealized stoichiometric crystal structure, without \ce{Pb} mix-in, found in natural mineral samples.
This chemical disorder is not expected to meaningfully perturb the dominant magnetic interactions.
Structurally, the \ce{Pb} ions reside at the hexagonal centers of the kagome lattice, spatially isolated from the active exchange pathways.
Furthermore, the electronic states associated with these intercalated ions lie deep below the Fermi level, preventing them from participating in the low-energy virtual hopping processes that govern superexchange. The idealized model serves as a robust approximation, and the magnitudes of our calculated exchange couplings should reflect the fundamental physics of the system.

\subsection{Magnetic ground state}

\begin{figure}
    \centering
    \includegraphics[width=1.0\linewidth]{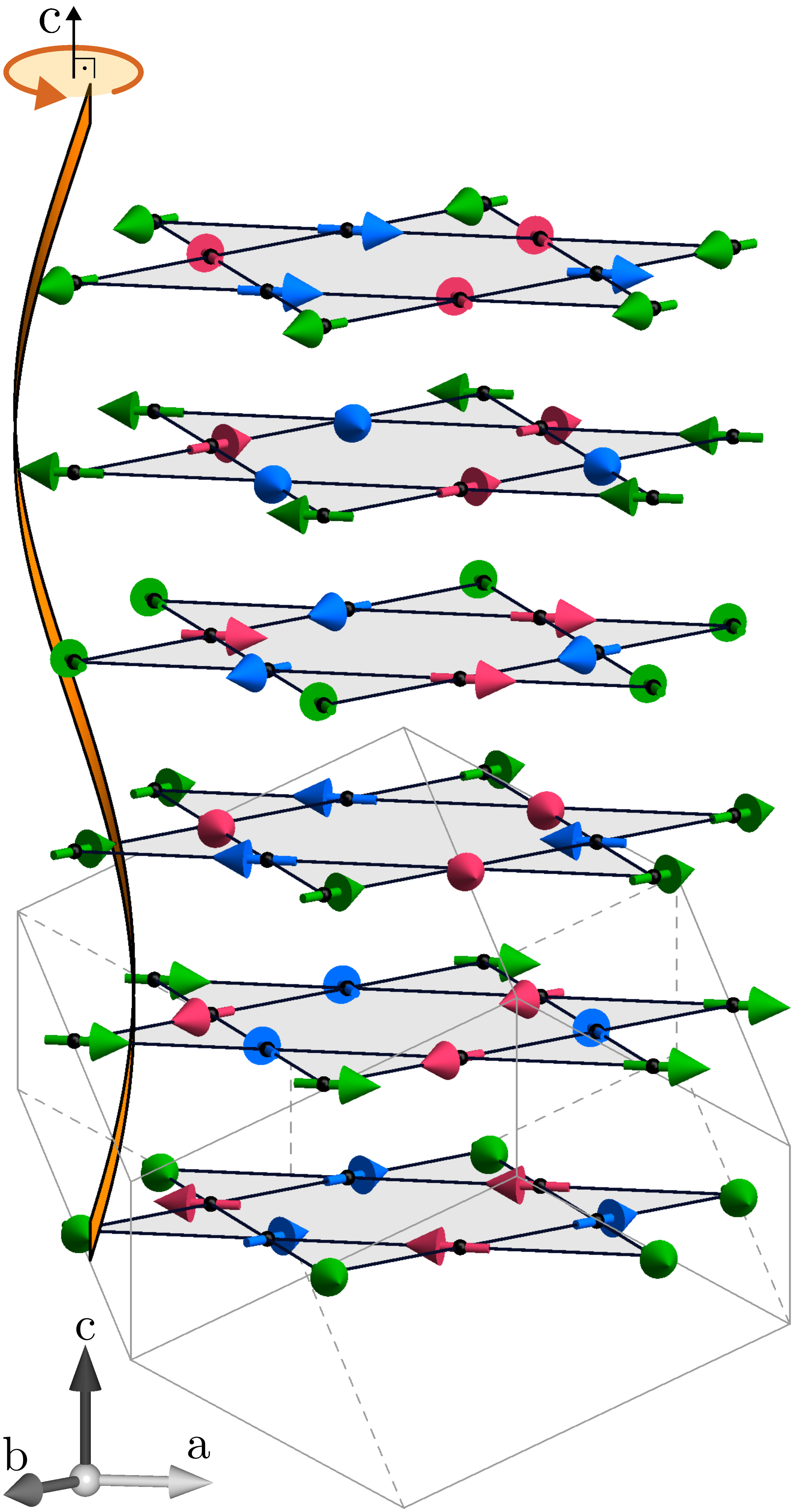}
    \caption{\textbf{Magnetic unit cell orderer.} Classical $\vec{Q}=(-2/3, -2/3, -1/6)$ state in real space. Each kagome layer forms a $\sqrt{3} \times \sqrt{3}$ magnetic structure. Adjacent layers, along the $c$-axis, feature a $60^\circ$ spin rotation, only in the right-hand orientation, highlighted by the orange band. The gray lines indicates the Wigner-Seitz magnetic unit cell, which needs only two layers.}
\label{fig:magnetic_order_chiarity}
\end{figure}

We investigate the ground state solution of the classical model with exchanges as listed in~\cref{tab:table_jnetwork_quatzo}.
 Before evaluating the full model including weaker interactions, we first establish a baseline by examining the reduced $J_2-J_3$ model restricted to the dominant couplings.
Using an iterative minimization approach, see \hyperref[sec:methods]{Methods}, we find that the extensive classical degeneracy of the isolated $J_2$ kagome network is lifted by the interlayer $J_3$ coupling, stabilizing a coplanar magnetic order, characterized by the propagation vector $\vec{Q}=(-2/3, -2/3, -1/6)$---expressed in terms of the crystal unit cell---or its counterparts related by three-fold rotational symmetry. In~\cref{fig:magnetic_order_chiarity} we show the magnetic unit cell which involves 18 sites, and display how it manifests on successive kagome layers. Each layer forms the well-known $\sqrt{3} \times \sqrt{3}$ magnetic order. While the propagation vector defines a particular direction at an obtuse angle with the $a$-$b$ plane, needing only 2 kagome layers to be written out, the perpendicular $c$-axis provides a much cleaner visual of the detailed structure: adjacent layers exhibit a $60^\circ$ spin rotation, forming a right-handed magnetic spiral.

The inherent chirality backed into the $60^\circ$-only right-handed rotations between layers, emerges from the symmetry considerations discussed earlier about distinguishing $J_3$ and $J_4$. If both bonds where symmetry equivalent, then there would be no right- or left-handed preference, as the layers would be coupled completely isotopically. However, removing  $J_4$ and considering only the $J_3$ bonds, we can see that the bond structure between adjacent triangles acquires a unique right-handed helical nature, see \cref{fig:j-couplingscombined}. This is also manifestly seen in the explicit construction of this ordered state, as detailed in the Supplemental Information.

Returning to the full model, and incorporating the next strongest exchange, $J_1$ elongates the spin spiral period, modifying the magnetic ordering vector along the $c$-axis, while the $\sqrt{3} \times \sqrt{3}$ order within the kagome planes, and the right-handed chirality of the spiral, remain unaltered.
The subsequent coupling $J_4$ exerts a similar but weaker effect on the magnetic order.
Finally, the smallest coupling $J_5$ does not influence the magnetic ground state.
Although AFM $J_5$ intrinsically competes with the in-plane $\sqrt{3} \times \sqrt{3}$ order, its magnitude is negligible compared to the dominant $J_2$ and $J_3$ couplings that enforce it.
Ultimately, the magnetic ordering vector of the full model with our specific exchange parameters is incommensurate and given by $\vec{Q} \approx (-2/3,-2/3,-0.16279)$, which we also confirm with the Luttinger Tisza method, see \hyperref[sec:methods]{Methods}.
This corresponds to a marginal 0.4\% shift in the $Q_z$ component compared to the reduced $J_2$-$J_3$ model. Note that the inclusion of spin-lattice coupling might circumvent the incommensurate order and lock $q_z$ to the commensurate $1/6$.

\subsection{Static and dynamic signatures}
\begin{figure}
	\centering
	\begin{overpic}[width=1.0\linewidth,percent,grid=false,tics=4]{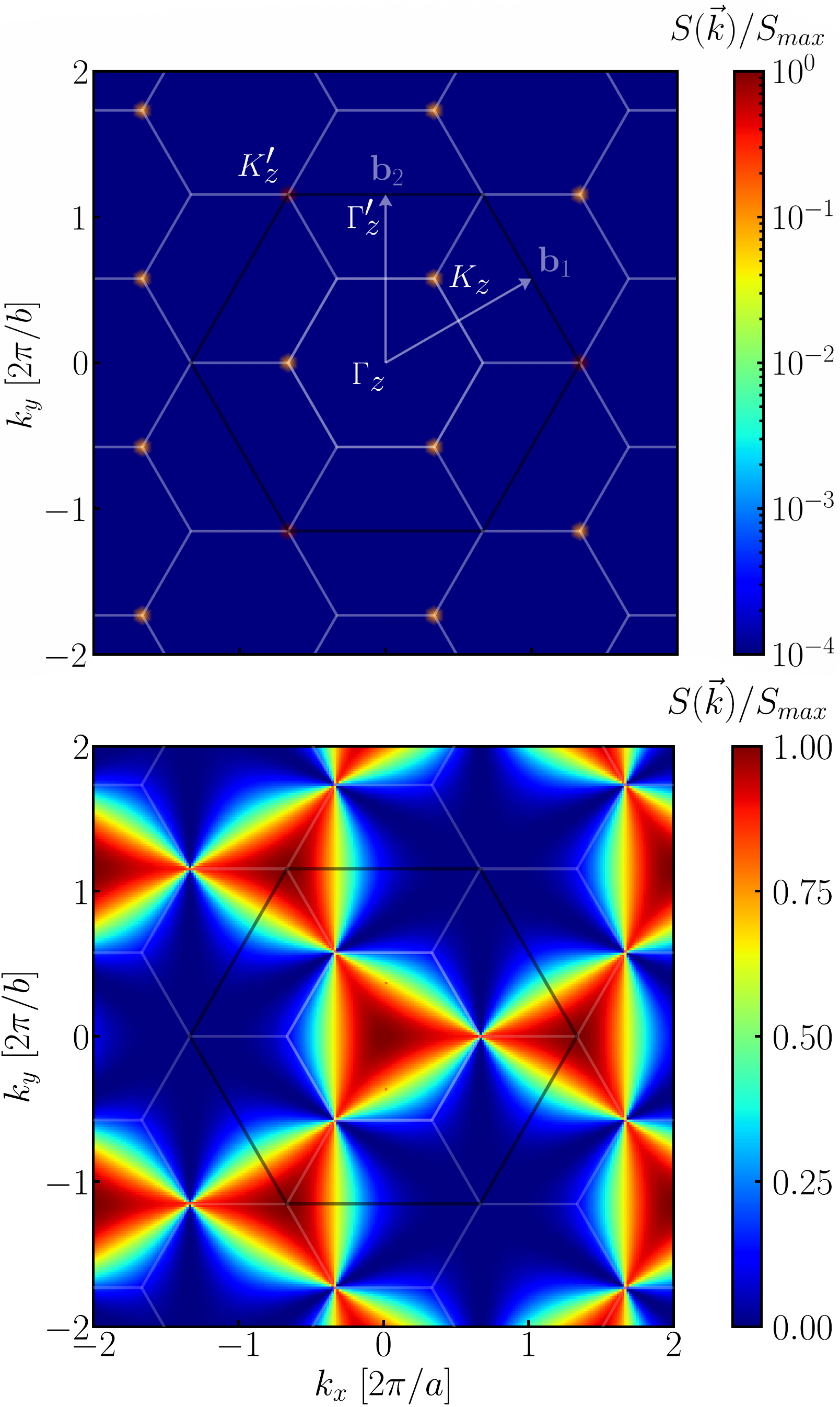}
        \put(0,98){\sffamily\textbf{a}}
        \put(0,50){\sffamily\textbf{b}}
	\end{overpic}
	\caption{\textbf{Static spin structure factor.} \textbf{a} The $q_z \approx -0.16279$ cut of $S(\vec{q})$ calculated for the full model \cref{tab:table_jnetwork_quatzo}. Solid white lines and black dashed lines indicate the boundaries of the first and the extended Brillouin zone, respectively. A dominant Bragg peak emerges at the ordering vector $\vec{Q}_1 = (-2/3, -2/3, q_z)=K'_z$, and three-fold symmetry-equivalent momenta, alongside a weaker secondary peak at $\vec{Q}_2 = (1/3, 1/3, q_z)=K_z$. \textbf{b} Static structure factor integrated only over the flat mode at $q_z = 0$ showing the pinch-point singularities located at the $K_z$ points.}
	\label{fig:SSF}
\end{figure}

\begin{figure}
	\centering
	\begin{overpic}[width=1.0\linewidth,percent,grid=false,tics=4]{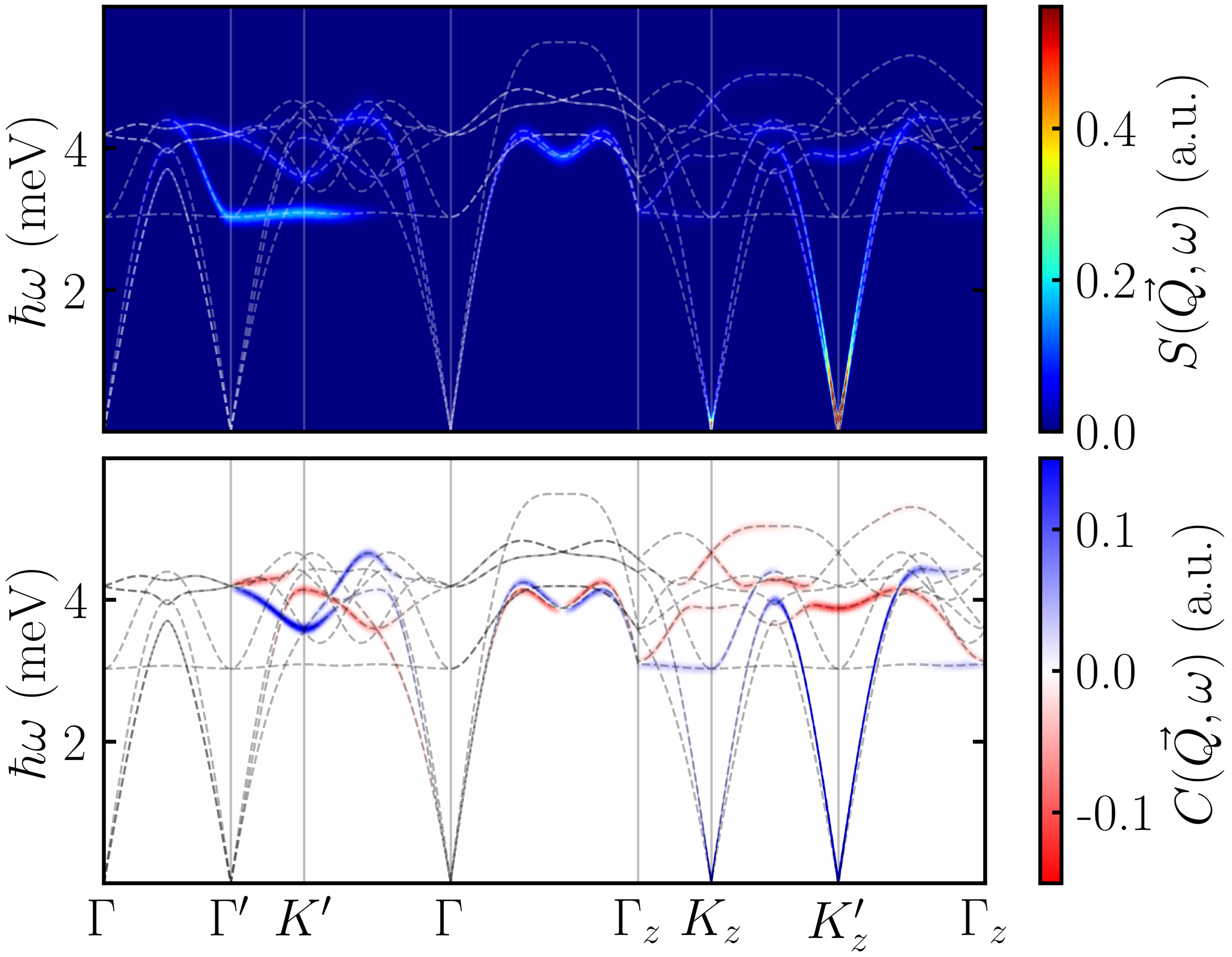}
		\put(0,77){\sffamily\textbf{a}}
		\put(0,40){\sffamily\textbf{b}}
	\end{overpic}
	\caption{\textbf{Magnon bands and dynamical factors.} Magnon excitation spectrum and dynamical chirality estimated via LSWT. The momentum path follows the high-symmetry points defined in \cref{fig:SSF}a, with $\Gamma' = (0,1,0)$, $K' = (-2/3,4/3,0)$, $\Gamma_z = (0,0,q_z)$, $K_z = (1/3,1/3,q_z)$, and $K'_z = (-2/3,4/3,q_z)$. \textbf{a} The magnon intensity strongly peaks at the primary magnetic ordering vectors $\vec{Q}_1=K_z'$ and $\vec{Q}_2=K_z$, in agreement with the classical structure factor. \textbf{b} Dynamical chirality signature, reflecting the right-handed spiral of the underlying magnetic ground state.
    The plots include Gaussian broadening of 0.1 meV. Note that the chirality operator changes sign in between $\Gamma$ and $\Gamma_z$, following the description in the main text.}
	\label{fig:quantzomagnons}
\end{figure}

Next, we evaluate the static and dynamic spin structure factors of the full model at zero temperature.
The static spin structure factor, shown in \cref{fig:SSF}a  exhibits a dominant Bragg peak at the magnetic ordering vector $\vec{Q}_1 =K'_z \approx (-2/3, -2/3, q_z) $ with $q_z \approx -0.16279$, see \cref{fig:quantzomagnons} , and three-fold rotations therein.
A secondary peak emerges at $\vec{Q}_2 = K_z \approx (1/3, 1/3, q_z)$, which carries approximately 10\% of the spectral intensity relative to the dominant peak.

The dynamic spin structure factors is calculated from LSWT, see \hyperref[sec:methods]{Methods}, with results in \cref{fig:quantzomagnons}a.
The spectral weight of the lowest-energy magnon modes is maximized at the gapless ordering wavevectors $\vec{Q}_1$ and $\vec{Q}_2$, consistent with the static structure factor signatures.
The magnon dispersion shows remarkably flat bands at an energy of roughly $3$~meV.
This lack of dispersion naturally produces a sharp peak in the magnon DOS.
Such a pronounced dispersionless feature serves as a fingerprint for the magnetic exchange model and could be directly probed in  momentum-integrated inelastic neutron scattering or Raman spectroscopy experiments.
By analyzing the magnon bands in different limits of the $J_3/J_2$ ratio, we find that the flat band originates from the zero-energy magnon modes of the AFM kagome lattice with only the first NN Heisenberg coupling.
Including the $J_3$ bonds into the model lifts the degeneracy of the zero-energy modes, pushing the magnon bands to higher energies, while one of the bands remains perfectly flat in the $k_x$--$k_y$ plane but is dispersive along the $k_z$ direction.
Further details on the magnon band structure and spin dynamics are provided in the Supplemental Information.

Further evidence for the connection to spin-liquid physics is provided by the static structure factor, where we integrate only over the flat mode of the reduced $J_2$-$J_3$ model, shown in \cref{fig:SSF}b.
In particular, the intensity exhibits characteristic pinch-point features suggestive of magnon fragmentation and indicative of local constraints and correlations associated with the spin-liquid regime~\cite{brooksbartlett2014magnetic,petit2016observation,essafi2016kagome,owen2016quantum,yan2018halfmoons,lhotel2020fragmentation,kiese2023pinch}.

These features therefore provide additional support for interpreting the flat magnon band as a remnant of the spin-liquid physics of the ideal kagome antiferromagnet.

The intrinsic chirality encoded in the right-handed spin-rotation structure of the ground state is also reflected in the system's dynamics. To characterize these dynamical chiral signatures, we calculate
%The inherent chirality backed into the right-hand spin rotation structure of the ground state can be further seen in the dynamics. We evaluate the dynamical chiral signatures of the system by calculating 
\begin{equation}
    C(\vec{Q},\omega) = S^{xy}(\vec{Q},\omega) - S^{yx}(\vec{Q},\omega).
\end{equation}
which is the antisymmetric part of the dynamical spin correlation tensor. For a collinear magnet with no chirality, $C(\vec{Q},\omega)$ is zero. In our case, however, the ground state is a right-handed spiral along the $c$-axis, so that we expect finite chirality magnon modes, carrying the same rotational character as the spiral order, as seen in \cref{fig:quantzomagnons}b.
Since the chirality originates from the layer-to-layer approximately $60^\circ$ spin rotation, the chirality signal is pronounced when scanning along reciprocal-space directions parallel to the spiral axis i.e. along $c$. 
We observe opposite chiral excitations along the $\Gamma-\Gamma_z$ line which flip the sign mid way, owing to sense of traversing the spiral in the positive and negative $c$-direction.
For this computation, we have restricted the coplanar magnetic state to the $a$-$b$ plane.
This choice of reference frame entails no loss of generality due to the underlying SU(2) symmetry of the Heisenberg exchanges.

In addition to the dynamic chirality $C(\vec{Q},\omega)$, the $\sqrt{3} \times \sqrt{3}$ magnetic order within each plane gives rise to a staggered vector chirality~\cite{grohol2005spin}.
However, due to the coplanar spin configuration, the scalar spin chirality, $\vec{S}_i \cdot (\vec{S}_j \times \vec{S}_k)$, vanishes.

\section{Discussion}

We have performed an extensive \textit{ab initio} study of quetzalcoatlite in its stoichiometrically idealized structure \ce{Zn6Cu3(TeO6)2(OH)6AgCl} and derived a minimal magnetic model for this kagome mineral. The electronic structure is governed by localized half-filled Cu $d_{x^2-y^2}$ orbitals, while the magnetic properties are dominated by two exchange interactions: a nearest-neighbor intralayer kagome coupling and a next-nearest-neighbor interlayer coupling. Looking at the microscopics of the ligand environment and exchange paths, we have argued that the exchange model found should be a good description even under \ce{Pb} mix-in. These interactions are sufficient to capture the essential features of the magnetic ground state.

Our classical analysis predicts a three-dimensionally ordered chiral magnetic phase in which each kagome layer exhibits a $\sqrt{3}\times\sqrt{3}$ order, while adjacent layers are rotated by $60^\circ$, forming a right-handed magnetic spiral along the crystallographic $c$-axis. Quetzalcoatlite therefore represents a rare realization of intrinsic chiral magnetism on a geometrically perfect kagome lattice.

The combination of an ideal kagome geometry, the absence of Cu/Zn antisite disorder, and comparatively small magnetic energy scales makes quetzalcoatlite a particularly promising platform for investigating frustrated and chiral magnetism. Moreover, the weak exchange couplings suggest tunability.
In particular, $J_3$ is responsible for the stable $\sqrt{3}\times\sqrt{3}$ order within the kagome planes.
%Furthermore, $J_5$ (second NN kagome layer exchange) is found to be exceedingly small, and is responsible for the presence of order.
Moderate pressure, chemical substitution, or structural modifications may significantly alter the balance of this exchange, potentially suppressing the chirality or the ordered state and promoting more strongly frustrated or even quantum spin-liquid phases.
Furthermore, $J_5$ (second NN kagome layer exchange) is found to be exceedingly small.
Tuning this coupling towards a stronger AFM or FM regime can destabilize or stabilize the coplanar spin order within the kagome planes. 
Quetzalcoatlite is a highly compelling candidate for future experimental synthesis and detailed characterization.

\section{Methods}\label{sec:methods}

\subsection{Structure relaxation}

Density functional theory calculations are performed in the \textit{Vienna ab-initio simulation package} (VASP)~\cite{kresse1993,kresse1996}, using the projector augmented-wave (PAW) method~\cite{Bloechl1994,kresse1999}.
We utilize the standard supplied PAW pseudopotentials  and consider the Perdew-Burke-Ernzerhof (PBE) ~\cite{perdew1996generalized} version of the generalized gradient approximation (GGA) as the exchange-correlation functional.
The plane-wave basis set cutoff energy is set to 600~eV, with an electronic energy convergence criterion of $10^{-7}$~eV. Brillouin zone integration is performed using a $6 \times 6 \times 6$ $\vec{k}$-point mesh.

As the starting point for our structural relaxation, we utilize the experimental crystallographic data for quetzalcoatlite, \ce{Zn_6Cu_3(TeO_6)_2(OH)_6 \cdot (Ag_xPb_yCl_{x+2y})}, reported in Ref.~\cite{burns2000quetzalcoatlite}, assuming no Pb mixing with Ag ($x=1$, $y=0$). 
Since the position of the hydrogen in the hydroxyl \ce{OH^-} groups were not resolved, we prepared 20 different realizations and performed constrained relaxations where only the hydrogen positions are updated, with a force convergence criterion of $10^{-3}$~eV/\AA, keeping the lattice vectors, unit cell shape, volume, and other species positions fixed.

\subsection{Electronic properties and TEMA}

The electronic band structure and DOS of quetzalcoatlite are calculated using the \textit{full-potential local-orbital} (FPLO) code package~\cite{fplo,fplo2}. We employ the PBE exchange-correlation functional~\cite{perdew1996generalized} with a reciprocal-space sampling of a $10 \times 10 \times 18$ $\vec{k}$-point mesh. On-site electron–electron correlations are treated within the GGA$+U$ framework in the atomic limit~\cite{czyzyk1994local}, using an on-site Coulomb repulsion $U = 6$~eV and a Hund’s exchange parameter $J_H = 1$~eV. The electronic self-consistency cycle is converged to a charge density tolerance of $10^{-6}$~eV.

To extract the magnetic exchange couplings, we perform a total energy mapping analysis (TEMA)~\cite{GlasbrennerNature2015,razpopov2023j,garcia2026microscopic}. 
For this purpose, we construct a $1 \times 2 \times 3$ supercell of the primitive unit cell, and compute total energies for a set of magnetic configurations using VASP~\cite{kresse1993,kresse1996}. For the exchange-correlation functional we use the PBE ~\cite{perdew1996generalized} version.
A kinetic energy cutoff of 600~eV sets the size of the plane-wave basis.
The Brillouin zone integration is performed on $4\!\times\!2\!\times\!2$ $\Gamma$-centered $\mathbf{k}$-grids using the tetrahedron method with Bl\"{o}chl corrections \cite{BloechlTetra1994}, and an energy criteria of $10^{-6}$ is used for the self-consistent total energy convergence.
Strong correlation effects associated with the localized \ce{Cu} $3d$ orbitals are treated using the rotationally invariant GGA$+U$ approach of Dudarev~\cite{dudarev1998electron}, with an effective Hubbard parameter $U_{\mathrm{eff}} = 6$~eV. In total, 45 distinct magnetic configurations are considered.

\subsection{Iterative torque updates}
The classical spin ground state of quetzalcoatlite is found by utilizing an iterative torque update algorithm on clusters spanning 30 $\times$ 30 $\times$ 30 and 15 $\times$ 15 $\times$ 86 in terms of the primitive unit cell.
The system is initialized in a random configuration, followed by reorientation updates
\begin{equation}
	\vec{S}_i \rightarrow -\frac{\vec{B}_i}{|\vec{B}_i|}, \quad \text{where} \quad \vec{B}_i = \sum_j J_{ij} \vec{S}_j.
\end{equation}
for a randomly selected spin $\vec{S}_i$, where $\vec{B}_i$ is the local field generated by neighbors of $i$.
This procedure is repeated until the maximum torque of all sites is below a threshold of $\mathrm{Max}(|\vec{S}_i\times\vec{B}_i|)<10^{-10}$.
To mitigate the risk of finding local energy minima, the minimization is executed multiple times from distinct random initializations, and the configuration with the lowest final energy is retained.

\subsection{Luttinger-Tisza method}
We confirm the incommensurability of the ground state by utilizing the Luttinger-Tisza (LT) method \cite{LuttingerTisza1946,Lyons1960Method}. 
In the LT method, the model is minimized subject to a single global spin-length constraint, specifically the sum of all spin lengths.
If the solution obtained under the relaxed (weak) constraint also satisfies the full local spin-length constraints, it corresponds to the exact ground state of the model. Even when these stronger constraints are not fulfilled, the resulting solution still provides a variational lower bound on the true ground-state energy.
  We use the LT implementation in SpinW package~\cite{Toth2015}, and confirm the helical nature of the ground state, with propagation vector and energy very close to the iterative minimization method. 

\subsection{Linear spin-Wave theory}
The low-energy magnetic excitation spectra and the stability of the optimized classical ground states are evaluated using linear spin-wave theory (LSWT), a bosonic expansion around the classical order, truncated at bilinear terms. The magnon dispersion is calculated via the incommensurate single Q-method as implemented in the SpinW package~\cite{Toth2015}.

\section{Data availability}
The datasets generated and analyzed during the current study are publicly available under the following link \url{https://doi.org/10.25716/gude.1fnk-rq57}. 

\section{Code availability}
The crystal structure was plotted with VESTA \cite{VESTA} (Version 3.5.7). Electronic bands and DOS are determined with FPLO~\cite{fplo,fplo2} (version 22.00-62). Exchange couplings are extracted from VASP ~\cite{kresse1993,kresse1996} (version 6.3.0). Ground state analysis of the spin model, including LSWT and LT, is performed in SpinW~\cite{Toth2015} (version v3.0).

The custom TEMA code used to extract the Heisenberg Hamiltonian, together with the iterative minimization code, is publicly available at \url{https://gitlab.itp.uni-frankfurt.de/razpopov/tema}.
%Custom code and scripts used for the iterative torque updates are not publicly available but may be made available to qualified researchers on reasonable request from the corresponding author.
    
%The underlying code for this study [and training/validation datasets] is not publicly available but may be made available to qualified researchers on reasonable request from the corresponding author. The underlying code [and training/validation datasets] for this study is available in [repository name] and can be accessed via this link [insert persistent URL to code]. 

\section{Acknowledgements}
The authors thank Daniel Lozano-G\'{o}mez for fruitful discussions.

\section{Funding}
A.R., P.P.S. and R.V. thank the Deutsche Forschungsgemeinschaft (DFG, German Research Foundation) through the TRR 288 - 422213477 (project A05, B05) and project Nr. VA 117/23-1 — 509751747.
F. F. acknowledges support from the Engineering and Physical Sciences Research Council, Grant No. EP/X012239/1.
M.R.N. was supported by the Materials Sciences and Engineering Division, Basic Energy Sciences, Office of Science, US Dept.~of Energy.

\section{Author contributions}
R.V., F.F. and M.R.N. conceived and supervised the project. A.R. performed the \textit{ab initio} calculations and extracted the effective spin model. P.P.S. carried out the classical model calculations. All authors contributed to the interpretation of the results and the writing of the manuscript.

\section{Competing interests}
The authors declare no competing financial or non-financial interests.

\section{Additional information}

%\subsection{Supplementary information }
%The online version contains supplementary material available at \textit{[URL to be inserted].}

\subsection{Correspondence}
Requests for materials should be addressed to \href{mailto:razpopov@itp.uni-frankfurt.de}{Aleksandar Razpopov} and  \href{mailto:valenti@itp.uni-frankfurt.de}{Roser Valent\'i}. 

\bibliographystyle{naturemag}
\bibliography{jour_name_abbreviation.bib,bibliography.bib}

\clearpage
\includepdf[pages=1, angle=0]{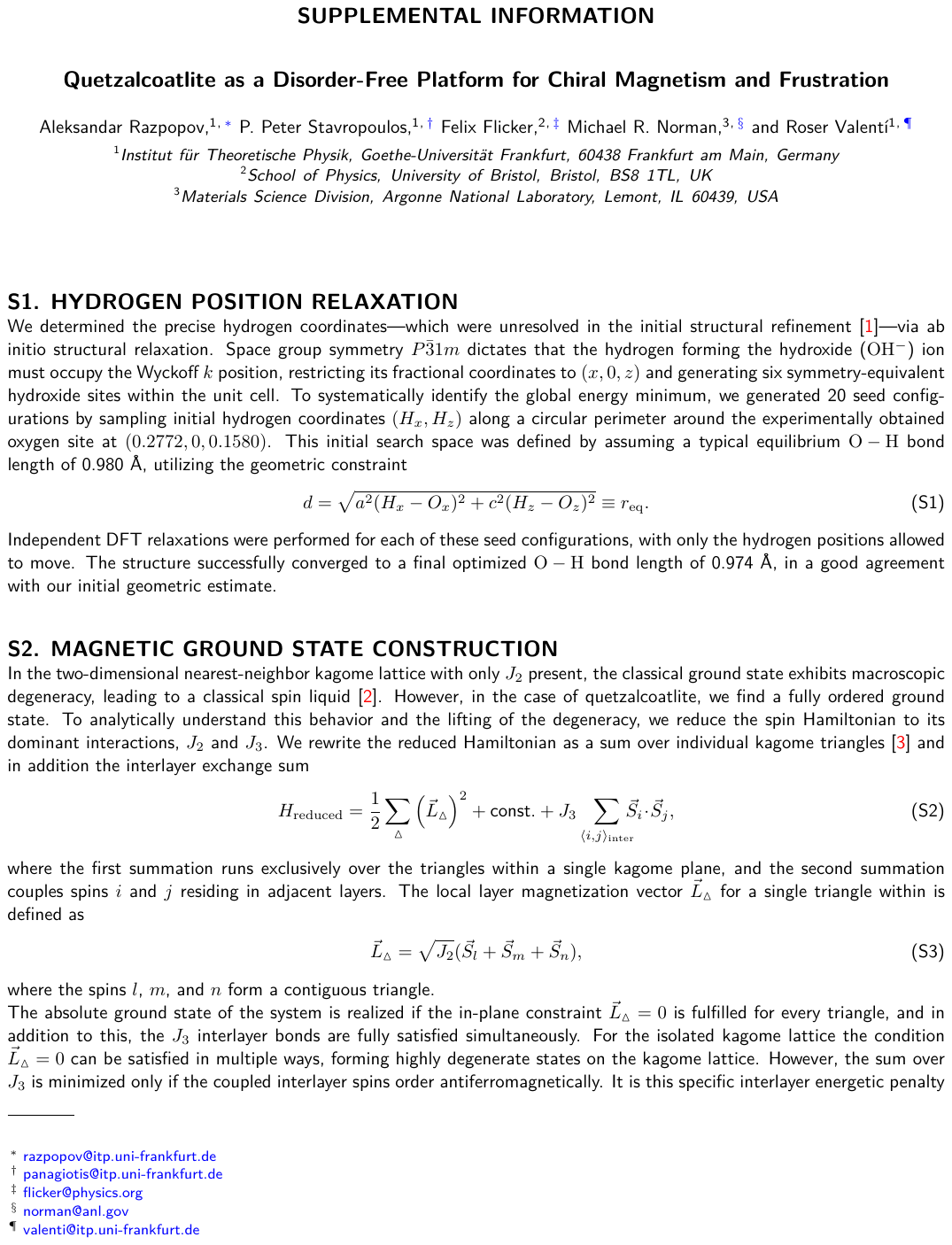}
\clearpage
\includepdf[pages=2, angle=0]{Supplement.pdf}
\clearpage
\includepdf[pages=3, angle=0]{Supplement.pdf}
\clearpage
\includepdf[pages=4, angle=0]{Supplement.pdf}
\clearpage
\includepdf[pages=5, angle=0]{Supplement.pdf}
\clearpage
\includepdf[pages=6, angle=0]{Supplement.pdf}

\end{document}